\documentclass[sn-nature, iicol]{sn-jnl}% Style for submissions to Nature Portfolio journals
\usepackage{graphicx}%
\usepackage{multirow}%
\usepackage{amsmath,amssymb,amsfonts}%
\usepackage{amsthm}%
\usepackage{mathrsfs}%
\usepackage[title]{appendix}%
\usepackage{xcolor}%
\usepackage{textcomp}%
\usepackage{manyfoot}%
\usepackage{booktabs}%
\usepackage{algorithm}%
\usepackage{algorithmicx}%
\usepackage{algpseudocode}%
\usepackage{listings}%
\usepackage{caption}
\theoremstyle{thmstyleone}%
\theoremstyle{thmstyletwo}%

\theoremstyle{thmstylethree}%

\renewcommand{\vec}[1]{\mathbf{#1}}

\newcounter{mycounter}

\begin{document}

\title[Article Title]{Breakdown of the Born-Oppenheimer approximation in solid hydrogen and hydrogen-rich solids}


\author*[1]{\fnm{Ville J.} \sur{H\"{a}rk\"{o}nen}}\email{ville.j.harkonen@gmail.com}



\affil*[1]{\orgdiv{Computational Physics Laboratory}, \orgname{Tampere University}, \orgaddress{\street{P.O. Box 692}, \postcode{FI-33014} \city{Tampere}, \country{Finland}}}
%\affil*[1]{\orgdiv{Department}, \orgname{Organization}, \orgaddress{\street{Street}, \city{City}, \postcode{100190}, \state{State}, \country{Finland}}}

%\affil[2]{\orgdiv{Department}, \orgname{Organization}, \orgaddress{\street{Street}, \city{City}, \postcode{10587}, \state{State}, \country{Country}}}

%\affil[3]{\orgdiv{Department}, \orgname{Organization}, \orgaddress{\street{Street}, \city{City}, \postcode{610101}, \state{State}, \country{Country}}}


\abstract{Hydrogen has been the subject of intense research following the discovery of high-temperature superconductivity in hydrides \cite{Drozdov-ConventionalSuperconductivityAt203KelvinAtHighPressuresInTheSulfurHydrideSystem-2015,Somayazulu-EvidenceForSuperconductivityAbove260KInLanthanumSuperhydrideAtMegabarPressures-PhysRevLett.122.027001-2019,Drozdov-SuperconductivityAt250KInLanthanumHydrideUnderHighPressures-2019}, and as a result of continuous efforts to produce solid hydrogen \cite{Dias-ObservationOfTheWignerHuntingtonTransitionToMetallicHydrogen-2017}. The Born-Oppenheimer approximation is the central piece of the quantum mechanical description of molecules and solids \cite{Born-Oppenheimer-Adiabatic-Approx.1927} and it is expected to have its weakest validity in hydrogen containing matter as it is the lightest element of all. The Born-Oppenheimer approximation is almost always assumed in the description of solids. Some beyond Born-Oppenheimer effects are likely included in the state-of-art method used to describe hydrogen-rich materials \cite{Errea-QuantumCrystalStructureInThe250KelvinSuperconductingLanthanumHydride-2020,Monacelli-QuantumPhaseDiagramOfHighPressureHydrogen-2023}, but the effects on the electronic structure in solids have not been considered before. Here we compute the beyond Born-Oppenheimer corrections \cite{Harkonen-ManyBodyGreensFunctionTheoryOfElectronsAndNucleiBeyondTheBornOppenheimerApproximation-PhysRevB.101.235153-2020} to electron density and report a breakdown of the Born-Oppenheimer approximation in experimentally known hydride superconductor YH$_{6}$ \cite{Troyan-AnomalousHighTemperatureSuperconductivityInYH6-2021} and in Cs-IV structure of solid hydrogen \cite{Tenney-PossibilityOfMetastableAtomicMetallicHydrogen-PhysRevB.102.224108-2020}. In both of these materials, we find a significant transfer of electron density from the volumes surrounding the expected positions of the hydrogen nuclei to volumes in between the nuclei. We expect these results to be the starting point of the beyond Born-Oppenheimer studies of electronic structure in solids, which is likely necessary to understand these forms of hydrogen-containing materials, also having significant technological importance.} %The abstract serves both as a general introduction to the topic and as a brief, non-technical summary of the main results and their implications. Authors are advised to check the author instructions for the journal they are submitting to for word limits and if structural elements like subheadings, citations, or equations are permitted.}




\maketitle

\section{Introduction}\label{sec1}

Hydrogen is the most common element in the universe and makes up about 75\% of all normal matter. Stars and large planets are usually made mostly of hydrogen \cite{Burbidge-SynthesisOfTheElementsInStars-RevModPhys.29.547-1957,Guillot-TheInteriorOfJupiter-2004,McMahon-ThePropertiesOfHydrogenAndHeliumUnderExtremeConditions-RevModPhys.84.1607-2012}. In recent years there has been a considerable scientific interest on physics related to hydrogen after the discovery of hydrogen rich superconductors \cite{Ashcroft-MetallicHydrogenAHighTemperatureSuperconductor-PhysRevLett.21.1748-1968,Drozdov-ConventionalSuperconductivityAt203KelvinAtHighPressuresInTheSulfurHydrideSystem-2015,Somayazulu-EvidenceForSuperconductivityAbove260KInLanthanumSuperhydrideAtMegabarPressures-PhysRevLett.122.027001-2019,Drozdov-SuperconductivityAt250KInLanthanumHydrideUnderHighPressures-2019}. Moreover, the solid hydrogen is the topic of active research of experimental \cite{Dias-ObservationOfTheWignerHuntingtonTransitionToMetallicHydrogen-2017} and theoretical science \cite{Wigner-OnThePossibilityOfaMetallicModificationOfHydrogen-1935,McMahon-GroundStateStructuresOfAtomicMetallicHydrogen-PhysRevLett.106.165302-2011,Tenney-PossibilityOfMetastableAtomicMetallicHydrogen-PhysRevB.102.224108-2020,Monacelli-QuantumPhaseDiagramOfHighPressureHydrogen-2023} as this information is relevant, for instance, in the study of planets \cite{McMahon-ThePropertiesOfHydrogenAndHeliumUnderExtremeConditions-RevModPhys.84.1607-2012}. The study of hydrogen thus holds significant importance both from a technological perspective and for enhancing our understanding of the universe.

Hydrogen has quite extreme properties related to its mass since it is the lightest of all elements making it difficult to describe in some respects. The Schr\"{o}dinger equation of hydrogen atom can be solved in an exact fashion given the Coulomb Hamiltonian for electron and proton, but for molecules and solids approximations are needed. The cornerstone of our current understanding of hydrogen in molecules and solids is the Born-Oppenheimer (BO) approximation \cite{Born-Oppenheimer-Adiabatic-Approx.1927,Born-Huang-DynamicalTheoryOfCrystalLattices-1954}. In the BO approximation the full Coulombic many-body problem is divided into two parts for electrons and for nuclei making the original problem computationally more feasible. The validity of the BO approximation relies on the large mass difference of electrons and nuclei. For this reason, we expect the BO approximation, in general, to have the weakest validity for hydrogen-containing matter.

Beyond-BO corrections in molecules have been studied extensively and the corrections are reported for numerous molecules and properties \cite{Schwenke-BeyondThePotentialEnergySurfaceAbInitioCorrectionsToTheBornOppenheimerApproximationForH2O-2001,Scherrer-OnTheMassOfAtomsInMoleculesBeyondTheBornOppenheimerApproximation-PhysRevX.7.031035-2017,Li-DFTofElectronTransferBeyondTheBornOppenheimerApproximationCaseStudyOfLiF-2018}. In solids, however, the beyond-BO corrections are rarely included. There are two exceptions to this rule. Firstly, the non-adiabatic phonon self-energy \cite{Calandra-AdiabaticAndNonadiabaticPhononDispersionInaWannierFunctionApproach-PhysRevB.82.165111-2010,Giustino-ElectronPhononInteractFromFirstPrinc-RevModPhys.89.015003-2017,Harkonen-ManyBodyGreensFunctionTheoryOfElectronsAndNucleiBeyondTheBornOppenheimerApproximation-PhysRevB.101.235153-2020} that is absent in the BO approximation. Also, the state-of-the-art computations on solid hydrogen \cite{Monacelli-QuantumPhaseDiagramOfHighPressureHydrogen-2023} and the recently discovered hydrogen-rich hydride compounds having a high-temperature superconductivity \cite{Errea-FirstPrinciplesTheoryOfAnharmonicityAndTheInverseIsotopeEffectInSuperconductingPalladiumHydrideCompounds-PhysRevLett.111.177002-2013,Errea-AnharmonicFreeEnergiesAndPhononDispersionsFromTheStochasticSCHarmonicApproximationApplicationToPlatinumAndPalladiumHydrides-PhysRevB.89.064302-2014,Errea-HighPressureHydrogenSulfideFromFirstPrinciplesAStronglyAnharmonicPhononMediatedSuperconductor-PhysRevLett.114.157004-2015,Errea-QuantumHydrogenBondSymmetrizationInTheSuperconductingHydrogenSulfideSystem-2016,Errea-QuantumCrystalStructureInThe250KelvinSuperconductingLanthanumHydride-2020} are using the stochastic self-consistent harmonic approximation (SSCHA) and is claimed to capture some beyond-BO effects. However, effects on electronic properties, like electron density, have not been considered earlier in solids.

In this work, we study the significance of beyond-BO corrections \cite{Harkonen-ManyBodyGreensFunctionTheoryOfElectronsAndNucleiBeyondTheBornOppenheimerApproximation-PhysRevB.101.235153-2020,Harkonen-ExactFactorizationOfTheManyBodyGreensFunctionTheoryOfElectronsAndNuclei-PhysRevB.106.205137-2022} to the electron density. We report a breakdown of the BO approximation by studying the electron density with quantum mechanical nuclei in two different structures at 0 K by using ab-initio computations based on density functional theory \cite{Hohenberg-DFT-PhysRev.136.B864-1964,KohnSham-DFT-PhysRev.140.A1133-1965}. We study the experimentally realized yttrium hydride YH$_{6}$ (at 150 GPa) \cite{Troyan-AnomalousHighTemperatureSuperconductivityInYH6-2021} and the Cs-IV phase of solid hydrogen (at 250 GPa) \cite{Tenney-PossibilityOfMetastableAtomicMetallicHydrogen-PhysRevB.102.224108-2020,Monacelli-QuantumPhaseDiagramOfHighPressureHydrogen-2023}. Both of these materials are stable at 0 K \cite{Troyan-AnomalousHighTemperatureSuperconductivityInYH6-2021,Tenney-PossibilityOfMetastableAtomicMetallicHydrogen-PhysRevB.102.224108-2020} with their respective pressures, within the BO harmonic approximation. We show that the beyond-BO corrections to the electron density are significant and beyond-BO methods are needed for the accurate description of materials like these.

%\begin{figure*}[h]%
%\centering
%\includegraphics[width=1.0\textwidth]{YH_fig1a_3.png}
%\caption{This is YH crystal structure.}
%\label{fig0}
%\end{figure*}
%\begin{figure}[h]%
%\centering
%\includegraphics[width=0.5\textwidth]{YH_fig1a_5.png}
%\caption{This is YH crystal structure.}
%\label{fig02}
%\end{figure}
\begin{figure}[h]%
\centering
\includegraphics[width=0.5\textwidth]{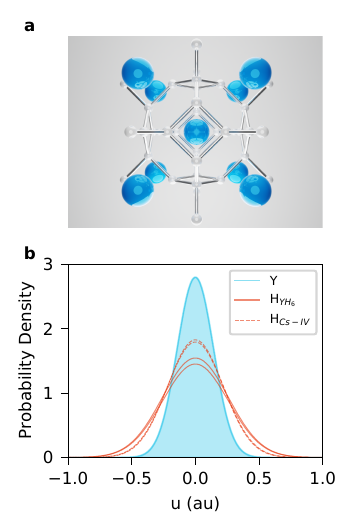}
\caption{\textbf{a} A conventional unit cell of the YH$_{6}$ crystal structure. \textbf{b} The one body nuclear probability densities, which are normal distributions for the ground state. There is only three different types of one body densities in total for YH$_{6}$ and two different densities for the hydrogen's in Cs-IV structure.}
\label{fig1}
\end{figure}
%\begin{figure}[h]%
%\centering
%\includegraphics[width=1.0\textwidth]{test_fig.png}
%\caption{Electron densities in 011 plane and in selected lines along the 001 plane.}
%\label{fig1}
%\end{figure}
\begin{figure*}[h]%
\centering
\includegraphics[width=1.0\textwidth]{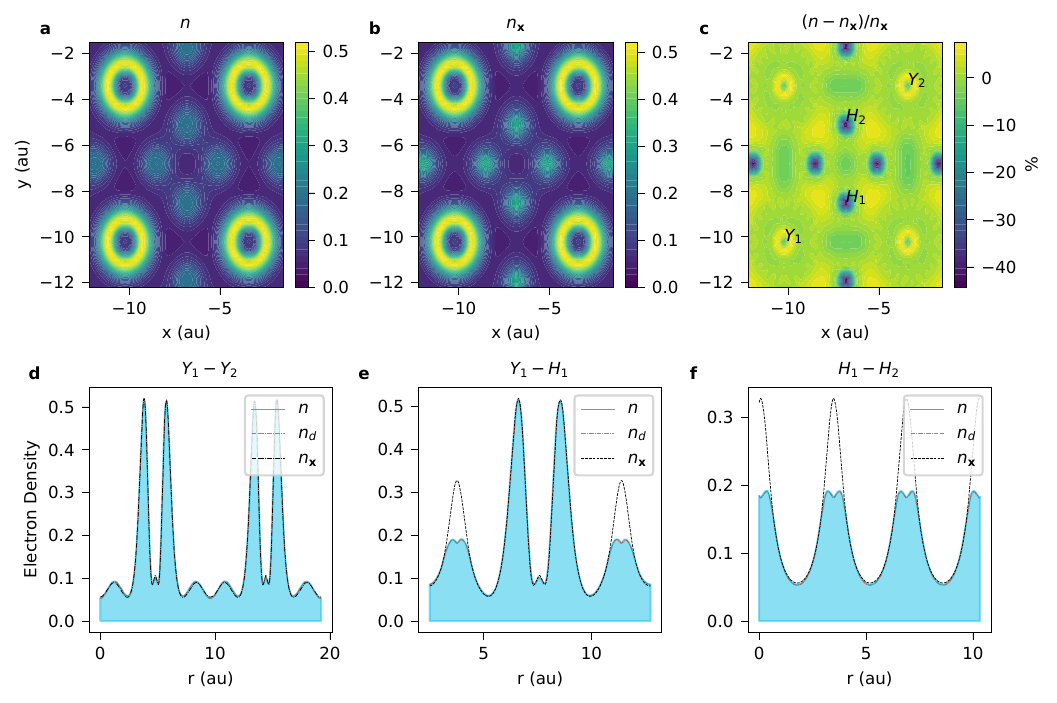}
\caption{Electron densities (valence electrons) in 11$\bar{1}$-plane of the primitive unit cell and in selected lines along the 11$\bar{1}$-plane. The electron densities are normalized to the number of (valence) electrons per unit cell. \textbf{a} The beyond-BO electron density $n\left(\vec{r}\right) = n_{\vec{x}}\left(\vec{r}\right) + n'_{\vec{x}}\left(\vec{r}\right)$, \textbf{b} the BO electron density $n_{\vec{x}}\left(\vec{r}\right)$, \textbf{c} the difference of beyond-BO and BO densities relative to $n_{\vec{x}}\left(\vec{r}\right)$ in percentage, \textbf{d} electron densities along the line between yttrium nuclei (pointed out in \textbf{c}), \textbf{e} between yttrium and hydrogen and \textbf{f} between hydrogen nuclei. In \textbf{e}-\textbf{f}, $n_{d}$ denotes the diagonal contribution to $n$ discussed in the text.}
\label{fig2}
\end{figure*}
\begin{figure*}[h]%
\centering
\includegraphics[width=1.0\textwidth]{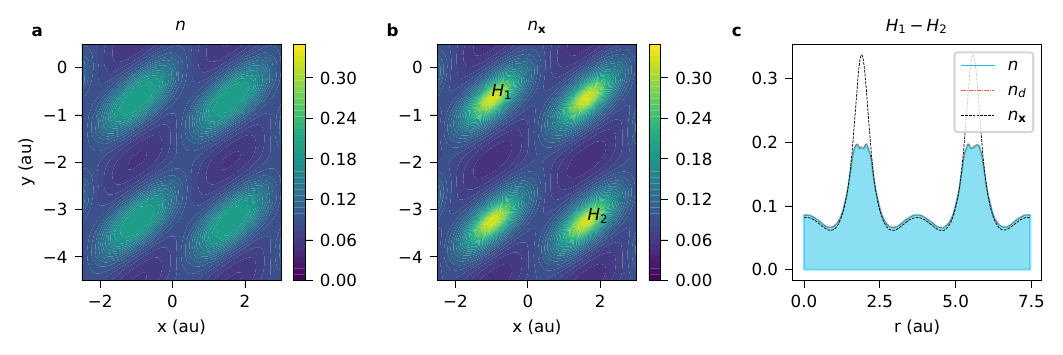}
\caption{Electron densities (valence electrons) in 111 plane and in selected lines along the 111 plane. \textbf{a} The beyond-BO electron density $n\left(\vec{r}\right) = n_{\vec{x}}\left(\vec{r}\right) + n'_{\vec{x}}\left(\vec{r}\right)$, \textbf{b} the BO electron density $n_{\vec{x}}\left(\vec{r}\right)$, \textbf{c} electron densities along the line connecting two hydrogen nucleus (pointed out in \textbf{b}).}
\label{fig3}
\end{figure*}
% The Introduction section, of referenced text expands on the background of the work (some overlap with the Abstract is acceptable). The introduction should not include subheadings.

\section{Results}
\label{Results}

The unit cell of the YH$_{6}$ crystal structure is depicted in Fig. \ref{fig1}\textbf{a}. This kind of visualization describes the essence of the BO approximation in the electronic problem. Namely, the nuclei with large masses have Dirac delta distributions and have well-defined positions. This is an approximation and the more rigorous picture is that the position of the nuclei are described by some probability density. The computed one-body nuclear densities for the vibrational ground state (normal distributed) are shown in Fig. \ref{fig1}\textbf{b}. Due to the symmetry, there are only few distinct one-body nuclear densities: one for yttrium atoms in YH$_{6}$ and two for hydrogen in both structures. The standard deviation for yttrium is about 0.14 au while for hydrogens it is 0.22-0.27 au, the highest values appearing in YH$_{6}$. The distance between the nearest hydrogen atoms are 2.41 au and 1.98 au for YH$_{6}$ and Cs-IV, respectively. This means that the equilibrium positions are about ten standard deviations apart and its unlikely that the nuclei would appear in the same region of space at low temperatures.

With the computed BO electronic and nuclear quantities, we can compute the corrections to the BO electron density. In Fig. \ref{fig2}, the computed BO and beyond-BO electron densities in the 11$\bar{1}$-plane are given. We see significant beyond-BO corrections to the BO electron density in the vicinity of hydrogen nuclei equilibrium positions, see Figs. \ref{fig2}\textbf{e} and \ref{fig2}\textbf{f}. The largest change of density, relative to the BO density, is around -45\%, see Fig. \ref{fig2}\textbf{c}. There is a decrease of electron density in the volumes surrounding the hydrogen nuclei and an increase of electron density in between yttrium nuclei, as well as in the surrounding volumes of the yttrium nuclei (see Figs. \ref{fig2}\textbf{c}-\ref{fig2}\textbf{e}). It is notable that the volume where the density is increased is much larger than the volume where the density is decreased (a stronger local change), and these are the volume sections surrounding the hydrogen nuclei. There is also a distinct change of the functional form of electron density at the hydrogen nuclei. Namely, the BO density at the location of hydrogen nuclei is unimodal while the beyond-BO density has a bimodal shape. We report very similar findings for the electron density in Cs-IV as shown in Fig. \ref{fig3}. A significant beyond-BO corrections can be seen at at the hydrogen nuclei equilibrium positions and the largest change of density relative to the BO density is about -44\%. The electron density thus increases in the space between the nuclei. The unimodal electron density shape is transformed to a bimodal shape when beyond-BO corrections are introduced.

A similar mechanism causes the beyond-BO corrections in both materials and can be deduced as follows. The results shown in Fig. \ref{fig1}\textbf{b} demonstrate that there is no significant overlap between probability densities of different nuclei. This in turn implies a local nature of the beyond-BO corrections. Namely, the uncertainty in the position of a particular nucleus is the main cause of the beyond-BO corrections at the equilibrium position of this nucleus and the other nuclei have only a secondary effect. This can be seen if we consider the beyond-BO corrections which can be be written as $n'_{\vec{x}}\left(\vec{r}\right) = \frac{1}{2} \sum_{s_{1}, s_{2}} \frac{\partial^{2}{n_{\vec{x}}\left(\vec{r}\right)}}{\partial{x_{s_{1}}} \partial{x_{s_{2}}}} \left(\boldsymbol{\Sigma}_{\vec{u}}\right)_{s_{1}s_{2}}$. In Figs. \ref{fig2}\textbf{d}-\ref{fig2}\textbf{f}, the different contributions to the beyond-BO corrections are separately plotted for YH$_{6}$. We see that essentially the whole correction originates from the diagonal elements of $n'_{\vec{x}}\left(\vec{r}\right)$ and in Figs. \ref{fig2}\textbf{d}-\ref{fig2}\textbf{f} corrections taking only these elements into account are denoted by $n_{d}$.  In the ground state, the covariance matrix elements are actually the expected values $\left(\boldsymbol{\Sigma}_{\vec{u}}\right)_{s_{1}s_{2}} = \left\langle 0|\hat{u}_{s_{1}} \hat{u}_{s_{2}}|0\right\rangle$. Thus we see that the diagonal elements must be positive and there are only two distinct diagonal elements in Cs-IV and three in YH$_{6}$. Since the beyond-BO corrections at the hydrogen equilibrium positions, $n'_{\vec{x}}\left(\vec{r}_{H}\right)$, are negative, we see that the second derivatives of electron density that are causing the corrections must be negative, $\frac{\partial^{2}{n_{\vec{x}}\left(\vec{r}_{H}\right)}}{\partial{x^{2}_{s}}} < 0$. The diagonal terms in the sum in the expression $n'_{\vec{x}}\left(\vec{r}_{H}\right)$ explain around 99\% of the beyond-BO corrections in YH$_{6}$ and Cs-IV. Moreover, only three of the diagonal terms in the sum explain around 94\% and 87\% of the beyond-BO corrections at $\vec{r}_{H}$ in YH$_{6}$ and Cs-IV, respectively.

The picture is thus rather simple from a physical point of view: the local position uncertainty of the hydrogen causes the change $n'_{\vec{x}}\left(\vec{r}_{H}\right)$ and only three terms out of $\left(3 N'_{n}\right)^{2}$ are essentially needed to explain the origin of these corrections. Similar reasoning applies to positive electron density corrections as supported by $n_{d}$ values plotted in Figs. \ref{fig2} and \ref{fig3}. The difference is that the electron density derivatives determining the sign of the corrections are positive in these volume sections.

\section{Conclusion}\label{sec13}

We have computed, for the first time, the beyond-BO corrections to electron density in the experimentally known high-temperature superconductor YH$_{6}$ \cite{Troyan-AnomalousHighTemperatureSuperconductivityInYH6-2021} and a predicted phase of solid hydrogen Cs-IV \cite{Tenney-PossibilityOfMetastableAtomicMetallicHydrogen-PhysRevB.102.224108-2020,Monacelli-QuantumPhaseDiagramOfHighPressureHydrogen-2023}. In both materials, we report significant deviations of electron densities from the values predicted by the BO approximation. Our findings suggest that in the studied materials, the beyond-BO corrections to electron density originate from the local uncertainty in nuclear positions and the correlation effects due to other nuclei seem to be insignificant. These results are reasonable from a physical point of view. Namely, in the electronic BO problem, the nuclei appear as infinite mass point particles located at their equilibrium positions. When nuclei are treated quantum mechanically, the location of the nuclei is more uncertain and the electrons, more or less, follow the nuclei \cite{Born-Huang-DynamicalTheoryOfCrystalLattices-1954}. Consequently, the beyond-BO electron density is more spread out away from the BO equilibrium positions as shown in this work.

We point out that the results presented in this work are expected to be valid only at low temperatures when the harmonic vibrational ground state describes the system with a reasonable accuracy. Even with this limitation, our approach can already be employed to enhance our understanding of the properties of hydrogen to support research focused on solid hydrogen production, given that the experiments are conducted at low temperatures \cite{Dias-ObservationOfTheWignerHuntingtonTransitionToMetallicHydrogen-2017}. At higher temperatures the situation becomes more complicated from a theoretical and computational point of view as the excited vibrational states have to be taken into account and also the anharmonic effects are expected to be more significant. Our theory \cite{Harkonen-ManyBodyGreensFunctionTheoryOfElectronsAndNucleiBeyondTheBornOppenheimerApproximation-PhysRevB.101.235153-2020,Harkonen-ExactFactorizationOfTheManyBodyGreensFunctionTheoryOfElectronsAndNuclei-PhysRevB.106.205137-2022} describes also these situations, but the complexity of the implementation will be more involved. It is also important to note that the exact solution of the Coulomb many-body electron-nuclei problem requires a self-consistent solution for the electronic and nuclear equations \cite{Harkonen-ManyBodyGreensFunctionTheoryOfElectronsAndNucleiBeyondTheBornOppenheimerApproximation-PhysRevB.101.235153-2020}. Here we computed the first iteration by using density functional theory and started the second iteration by computing the beyond-BO corrected electron density. The next step would be to update the nuclear reference positions and to use the computed nuclear quantities as a part of the electronic equations. Some of these effects are possibly included to the SSCHA approach that has been used to compute the corrected nuclear reference positions (due to the proton position uncertainty) in hydrides \cite{Errea-FirstPrinciplesTheoryOfAnharmonicityAndTheInverseIsotopeEffectInSuperconductingPalladiumHydrideCompounds-PhysRevLett.111.177002-2013,Errea-HighPressureHydrogenSulfideFromFirstPrinciplesAStronglyAnharmonicPhononMediatedSuperconductor-PhysRevLett.114.157004-2015,Errea-QuantumHydrogenBondSymmetrizationInTheSuperconductingHydrogenSulfideSystem-2016,Errea-QuantumCrystalStructureInThe250KelvinSuperconductingLanthanumHydride-2020} and solid hydrogen \cite{Monacelli-QuantumPhaseDiagramOfHighPressureHydrogen-2023}. Similar result has been reported earlier for ice at high pressures \cite{Benoit-TunnelingAndZeroPointMotionInHighPressureIce-1998}.

We see the breakdown of the BO approximation here two fold and possibly connected to these previous results. Firstly, if the total force on the nuclei vanish also with the beyond-BO electron density, then the equilibrium positions will be the same as within the BO approximation. In this case, the structures will be the same as in the BO approximation, but there is a rather large change in the electronic structure as illustrated by Figs. \ref{fig2} and \ref{fig3}. On the other hand, if the total force on each nucleus with the beyond-BO electron density does not vanish, then the crystal structure will change in one way or the other. This latter option might be the case in Refs. \cite{Errea-FirstPrinciplesTheoryOfAnharmonicityAndTheInverseIsotopeEffectInSuperconductingPalladiumHydrideCompounds-PhysRevLett.111.177002-2013,Errea-HighPressureHydrogenSulfideFromFirstPrinciplesAStronglyAnharmonicPhononMediatedSuperconductor-PhysRevLett.114.157004-2015,Errea-QuantumHydrogenBondSymmetrizationInTheSuperconductingHydrogenSulfideSystem-2016,Errea-QuantumCrystalStructureInThe250KelvinSuperconductingLanthanumHydride-2020,Benoit-TunnelingAndZeroPointMotionInHighPressureIce-1998} and a recent study \cite{Monacelli-QuantumPhaseDiagramOfHighPressureHydrogen-2023} suggests that this is also true for Cs-IV. That is, the BO harmonic approximation gives a apparently stable structures, but when beyond-BO corrections are included, the electronic structure is altered and this will likely cause a change to the equilibrium positions. In both of these outcomes, our results imply that the BO approximation is not valid and there likely are measurable consequences of its breakdown.

To summarize. We report a breakdown of the BO approximation in two hydrogen containing solids which manifest itself as a change in the electronic structure of these materials. To better understand solid hydrogen and hydrogen-rich materials, we call for the necessity for the inclusion of beyond-BO effects in their description. We believe that the results presented here will aid towards a deeper understanding of various phenomena related to hydrogen, like high-temperature superconductivity and even in the study of planets.

%Conclusions may be used to restate your hypothesis or research question, restate your major findings, explain the relevance and the added value of your work, highlight any limitations of your study, describe future directions for research and recommendations. 

%In some disciplines use of Discussion or 'Conclusion' is interchangeable. It is not mandatory to use both. Please refer to Journal-level guidance for any specific requirements. 

\bibliography{sn-bibliography}% common bib file

\backmatter

%\bmhead{Supplementary information}

%If your article has accompanying supplementary file/s please state so here. 

%Authors reporting data from electrophoretic gels and blots should supply the full unprocessed scans for key as part of their Supplementary information. This may be requested by the editorial team/s if it is missing.

%Please refer to Journal-level guidance for any specific requirements.

\bmhead{Acknowledgments}

The author acknowledges Prof. E. K. U. Gross for numerous discussions on beyond-BO physics over the years and Prof. Esa R\"{a}s\"{a}nen for comments on the manuscript. The author gratefully acknowledges funding from the Magnus Ehrnrooth foundation and Jenny and Antti Wihuri foundation. The computing resources for this work were provided CSC - the Finnish IT Center for Science.

%Acknowledgments are not compulsory. Where included they should be brief. Grant or contribution numbers may be acknowledged.
%Please refer to Journal-level guidance for any specific requirements.

\section*{Declarations}

%Some journals require declarations to be submitted in a standardised format. Please check the Instructions for Authors of the journal to which you are submitting to see if you need to complete this section. If yes, your manuscript must contain the following sections under the heading `Declarations':

\bmhead{Competing interests} The authors declare no competing interests.

\bmhead{Availability of data and materials} All the data generated in this work is available upon a request.

\bmhead{Code availability} Quantum Espresso is an open-access code available for public use. The codes used to compute the electron density corrections from Quantum Espresso output are private codes developed
by the author and are being prepared for distribution as an open-source code.

%\noindent
%If any of the sections are not relevant to your manuscript, please include the heading and write `Not applicable' for that section. 

%\bigskip

%\bigskip\noindent
%Springer journals and proceedings: \url{https://www.springer.com/gp/editorial-policies}

%\bigskip\noindent
%Nature Portfolio journals: \url{https://www.nature.com/nature-research/editorial-policies}

%\bigskip\noindent
%\textit{Scientific Reports}: \url{https://www.nature.com/srep/journal-policies/editorial-policies}

%\bigskip\noindent
%BMC journals: \url{https://www.biomedcentral.com/getpublished/editorial-policies}
%\end{flushleft}

\section{Methods}
\label{sec11}

%Topical subheadings are allowed. Authors must ensure that their Methods section includes adequate experimental and characterization data necessary for others in the field to reproduce their work. Authors are encouraged to include RIIDs where appropriate. 

\subsection{Electron density}
\label{ElectronDensity}

By combining the beyond-BO Green's function theory \cite{Harkonen-ManyBodyGreensFunctionTheoryOfElectronsAndNucleiBeyondTheBornOppenheimerApproximation-PhysRevB.101.235153-2020} with the exact factorization of the wave function \cite{Hunter-ConditionalProbInWaveMech-1975,Gidopoulos-ElectronicNonAdiabaticStates-2005,Gidopoulos-Gross-ElectronicNonAdiabaticStates-2014}, it can be shown \cite{Harkonen-ExactFactorizationOfTheManyBodyGreensFunctionTheoryOfElectronsAndNuclei-PhysRevB.106.205137-2022} that the zero temperature electronic Green's function can be written approximatively as $G\left(\vec{r}t,\vec{r}'t'\right) \approx \int d\vec{R} \left|\chi\left(\vec{R}\right)\right|^{2} G_{\vec{R}}\left(\vec{r}t,\vec{r}'t'\right)$ and thus for the electron density
\begin{equation} 
n\left(\vec{r}\right) = \int d\vec{R} \left|\chi\left(\vec{R}\right)\right|^{2} n_{\vec{R}}\left(\vec{r}\right).
\label{eq:ElectronDensityEq_1}
\end{equation}
Here, $\chi\left(\vec{R}\right)$ is the nuclear many-body wave function satisfying $H_{n} \chi\left(\vec{R}\right) = E \chi\left(\vec{R}\right)$ a special case of which is the BO nuclear equation \cite{Harkonen-ExactFactorizationOfTheManyBodyGreensFunctionTheoryOfElectronsAndNuclei-PhysRevB.106.205137-2022}. Moreover, $n_{\vec{R}}\left(\vec{r}t\right)$ is the BO electron density that can be obtained from the electronic BO equation (or from the electronic Green's function). The beyond-BO corrected electron density is thus the expected value of the BO electron density with respect to the nuclear density $\left|\chi\left(\vec{R}\right)\right|^{2}$. Here we concentrate to system at zero absolute temperature meaning that the vibrational system is in its ground state. In this case, the nuclei are often rather close to their equilibrium positions and the harmonic approximation is often justified. We thus assume that $H_{n}$ is the harmonic BO nuclear Hamiltonian and the solution of the corresponding Schr\"{o}dinger equation is known \cite{Born-Huang-DynamicalTheoryOfCrystalLattices-1954,Harkonen-OnTheDiagonalizationOfQuadraticHamiltonians-2021} in the normal coordinates $\vec{q}$ such that $\chi_{0}\left(\vec{q}\right) = \chi_{0}\left(\vec{q}_{1}\right) \cdots \chi_{0}\left(\vec{q}_{3N'_{n}}\right)$. Here each $\chi_{0}\left(\vec{q}_{j}\right)$ are of the simple harmonic oscillator form. The corresponding nuclear density $\left|\chi_{0}\left(\vec{q}\right)\right|^{2}$ is of the multivariate normal form with the covariance matrix $\boldsymbol{\Sigma}_{\vec{q}} = \text{diag}\left(\frac{\hbar}{2 \omega_{1}}, \ldots, \frac{\hbar}{2 \omega_{3N'_{n}}}\right)$, where $\omega_{j}$ are the normal mode frequencies. The normal coordinates $\vec{q}$ are connected to the displacements $\vec{u} = \vec{R} - \vec{x}$ from the equilibrium positions $\vec{x}$ by a linear transformation $\vec{u} = \vec{C} \vec{q}$. Thus, since $\vec{q}$ are normal distributed, also $\vec{R}$ are \cite{Tong-TheMultivariateNormalDistribution-1990}
\begin{equation} 
\left|\chi_{0}\left(\vec{R}\right)\right|^{2} = \frac{ \exp\left[-\frac{1}{2} \vec{u}^{T} \boldsymbol{\Sigma}^{-1}_{\vec{u}} \vec{u}\right] }{\sqrt{ \left(2 \pi\right)^{3N_{n}} \left|\boldsymbol{\Sigma}_{\vec{u}} \right|}},
\label{eq:ElectronDensityEq_2}
\end{equation}
where the covariance matrix is $\boldsymbol{\Sigma}_{\vec{u}} = \vec{C} \boldsymbol{\Sigma}_{\vec{q}} \vec{C}^{T} = \vec{M}^{-1} \vec{e} \boldsymbol{\Sigma}_{\vec{q}} \vec{e}^{T} \vec{M}^{-1}$. Here $\vec{M}$ is matrix with nuclear masses in its diagonal and $\vec{e}$ are the eigenvectors of the mass scaled interatomic force constant matrix, which is the second order mixed partial derivative of the BO energy with respect to the nuclear equilibrium positions $\vec{x}$. We can now compute the beyond-BO corrected electron density by using Eqs. \eqref{eq:ElectronDensityEq_1} and \eqref{eq:ElectronDensityEq_2} with the quantities obtainable from BO calculations. The computational problem originates, however, from the complexity of the BO electron density $n_{\vec{R}}\left(\vec{r}\right)$ as a function of the nuclear variables $\vec{R}$. We are considering here the zero temperature situation in which the expected values of the nuclear mean square displacements are the smallest \cite{Harkonen-NTE-2014}. We thus expand $n_{\vec{R}}\left(\vec{r}\right) = n_{\vec{x} + \vec{u}}\left(\vec{r}\right)$ to a Taylor series in $\vec{u}$ about $\vec{x}$ and up to second order, we approximate Eq. \eqref{eq:ElectronDensityEq_1} for the vibrational ground state
\begin{equation}
n\left(\vec{r}\right) \approx n_{\vec{x}}\left(\vec{r}\right) + n'_{\vec{x}}\left(\vec{r}\right),
\label{eq:ElectronDensityEq_3}
\end{equation}
where the second-order beyond-BO correction is
\begin{equation}
n'_{\vec{x}}\left(\vec{r}\right) \equiv \frac{1}{2} \sum_{s_{1}, s_{2}} \frac{\partial^{2}{n_{\vec{x}}\left(\vec{r}\right)}}{\partial{x_{s_{1}}} \partial{x_{s_{2}}}} \left(\boldsymbol{\Sigma}_{\vec{u}}\right)_{s_{1}s_{2}}.
\label{eq:ElectronDensityEq_4}
\end{equation}
The quantities needed to compute $n\left(\vec{r}\right)$ are therefore the equilibrium BO electron density $n_{\vec{x}}\left(\vec{r}\right)$, its second order mixed partial derivatives and the covariance matrix $\boldsymbol{\Sigma}_{\vec{u}}$. All these quantities within the BO approximation can be obtained by using many available open source ab-initio computational packages. We give calculational details in Sec. \ref{CalculationalDetails}.

\subsection{Center-of-mass frame}
\label{CenterOfMassFrame}

There is a subtle issue that needs to be addressed when computing the densities from Eq. \eqref{eq:ElectronDensityEq_3}. This topic has been extensively discussed in the literature \cite{Sutcliffe-TheDecouplingOfElectronicAndNuclearMotions-2000,Kreibich-MulticompDFTForElectronsAndNuclei-PhysRevLett.86.2984-2001,vanLeeuwen-FirstPrincElectronPhonon-PhysRevB.69.115110-2004,Kreibich-MulticompDFTForElectronsAndNuclei-PhysRevA.78.022501-2008,Harkonen-ManyBodyGreensFunctionTheoryOfElectronsAndNucleiBeyondTheBornOppenheimerApproximation-PhysRevB.101.235153-2020,Harkonen-ExactFactorizationOfTheManyBodyGreensFunctionTheoryOfElectronsAndNuclei-PhysRevB.106.205137-2022}, but is usually absent in the literature regarding lattice dynamics, which was originally formulated in the laboratory frame of reference \cite{Born-Huang-DynamicalTheoryOfCrystalLattices-1954}. The so-called acoustic sum rule, that originates from the translational symmetry \cite{Born-Huang-DynamicalTheoryOfCrystalLattices-1954}, dictates that there are three vibrational modes of zero frequency corresponding to the translational motion of the system as a whole. For this reason the covariance matrix $\boldsymbol{\Sigma}_{\vec{q}}$ discussed in Sec. \ref{ElectronDensity} is not invertible and thus the density $\left|\chi_{0}\left(\vec{R}\right)\right|^{2}$ is not well defined in the laboratory frame. For this reason we denoted in Sec. \ref{ElectronDensity} by $N'_{n} = N_{n} - 3$ the number of nuclei in the nuclear center-of-mass frame,  $N_{n}$ the number of nuclei in the laboratory frame and $\vec{R}$ denotes the nuclei coordinates in the nuclear center-of-mass frame. To resolve this issue \cite{Harkonen-ManyBodyGreensFunctionTheoryOfElectronsAndNucleiBeyondTheBornOppenheimerApproximation-PhysRevB.101.235153-2020}, we separate the motion of the nuclear center of mass from the internal motion of the system and the resulting harmonic nuclear Hamiltonian becomes
\begin{equation}
H_{n} = \sum^{N_{n}-1}_{k = 1} \frac{ \vec{p}^{2}_{k}}{2 M_{k}} + \Phi'_{2},
\label{eq:CenterOfMassFrameEq_1}
\end{equation}
where the harmonic potential energy is
\begin{align}
\Phi'_{2} =& \frac{1}{2} \sum^{N_{n}-1}_{k'_{1}, k'_{2} = 1} \sum_{\alpha_{1}, \alpha_{2} } \Phi'_{\alpha_{1}\alpha_{2}}\left(k'_{1}, k'_{2}\right) \nonumber \\
          &\times u_{\alpha_{1}}\left(k'_{1}\right) u_{\alpha_{2}}\left(k'_{2}\right).
\label{eq:CenterOfMassFrameEq_2}
\end{align}
Moreover, the interatomic force constants $\Phi'_{\alpha_{1}\alpha_{2}}\left(k'_{1}, k'_{2}\right)$ in the center-of-mass frame are connected to the laboratory frame force constants, $\Phi_{\alpha_{1}\alpha_{2}}\left(k_{1}, k_{2}\right)$, as
\begin{align}
\Phi'_{\alpha_{1} \alpha_{2}}\left(k'_{1}, k'_{2}\right) =& \sum^{N_{n}}_{k_{1}, k_{2} = 1} \Phi_{\alpha_{1} \alpha_{2}}\left(k_{1}, k_{2}\right) \nonumber \\
&M'_{k_{1}k'_{1}} M'_{k_{2}k'_{2}}.
\label{eq:CenterOfMassFrameEq_3}
\end{align}
where $M'_{k_{j}k'_{j}} \equiv \delta_{k'_{j}k_{j}} - \delta_{k_{j}N_{n}} M_{k'_{j}} / M_{N_{n}}$. In the electronic BO equation, the nuclear variables $\vec{R}$ can be treated as parameters and thus after transforming the electronic coordinates to the nuclear center-of-mass frame we have an identity transformation provided we choose the nuclear center-of-mass to be at the origin. With this choice $n_{\vec{x}}\left(\vec{r}\right)$ is the same for both frames of reference. We can now solve the nuclear problem with the Hamiltonian given by Eq. \eqref{eq:CenterOfMassFrameEq_1} by using the same approaches as in the laboratory frame formulation \cite{Born-Huang-DynamicalTheoryOfCrystalLattices-1954} and thus compute the corrections to electron density by using Eqs. \eqref{eq:ElectronDensityEq_1} and \eqref{eq:ElectronDensityEq_3}.

\subsection{Calculational details}
\label{CalculationalDetails}

All ab-initio computations of this work were done by using Quantum Espresso (QE) program package \cite{Giannozzi-QuantumEspresso-2009} (version 7.0). We use PBE functional \cite{Perdew-GeneralizedGradientApproximationMadeSimple-PhysRevLett.77.3865-1996} and GBRV pseudopotentials \cite{Garrity-PseudopotentialsForHighThroughputDFTCalculations-2014} (version 1.4) for hydrogen and yttrium. The harmonic phonon frequencies are computed by using the density functional perturbation theory as implemented in QE \cite{Baroni-PhononsAndRelatedCrystalPropFromDFTPT-RevModPhys.73.515-2001}. The plane wave kinetic energy and charge density cut-off values used were 90 Ry and 360, respectively. In all computations, we use Gaussian smearing of 0.001 Ry. The electronic structure was computed with $12 \times 12 \times 12$ (YH$_{6}$) and $32 \times 32 \times 16$ (Cs-IV) $\vec{k}$ point grids. We constructed $2 \times 2 \times 2$ supercells in order to compute the electron density derivatives of Eq. \eqref{eq:ElectronDensityEq_4}. The derivatives were computed as finite central differences with 0.5\% displacements from the nuclear equilibrium positions. To compute the electron density corrections from Eq. \eqref{eq:ElectronDensityEq_3} we computed the lattice dynamical properties (ph.x module of QE) with the $\vec{q}$ point meshes matching the supercell dimensions.

The structures used, YH$_{6}$ ($Im\bar{3}m$) and Cs-IV ($I4_{1}/amd$), can be found from the supplementary materials of \cite{Tenney-PossibilityOfMetastableAtomicMetallicHydrogen-PhysRevB.102.224108-2020,Troyan-AnomalousHighTemperatureSuperconductivityInYH6-2021}. The YH$_{6}$ ($Im\bar{3}m$) structure parameters used are the following: lattice parameter $a = 3.602$ \r{A}; fractional coordinates of the in-equivalent atoms $Y_{1} = \left(0.000, 0.000, 0.000\right)$, $H_{1} = \left(0.250, 0.000, 0.500\right)$. The Cs-IV ($I4_{1}/amd$) structure parameters: lattice parameters $a = 1.37369$ \r{A}, $c = 3.15838$ \r{A}; fractional coordinates of the in-equivalent atoms $H_{1} = \left(0.000, 0.750, 0.125\right)$ and $H_{2} = \left(0.000, 0.250, 0.875\right)$. We first established the structure relaxation of these structures with the given parameters after which the lattice dynamical properties were computed.

\newpage

\begin{figure*}[h]%
\centering
\includegraphics[width=1.0\textwidth]{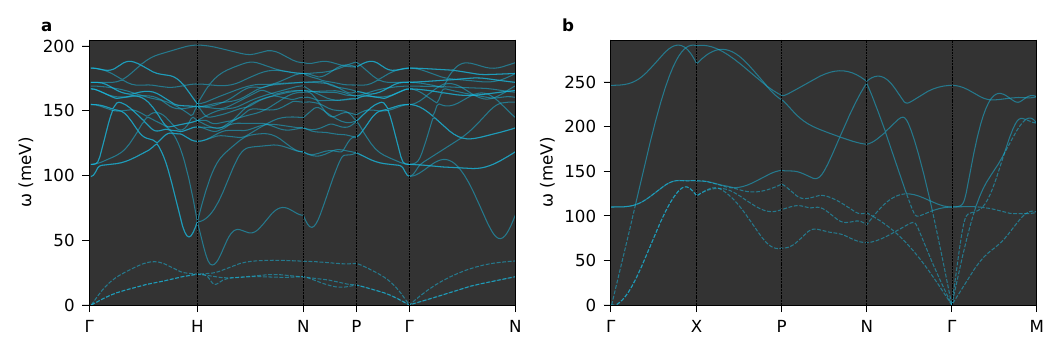}
\stepcounter{mycounter}
\caption*{\textbf{Extended Data Fig. \arabic{mycounter}} The computed harmonic phonon dispersion relations within the BO approximation for \textbf{a} YH$_{6}$, \textbf{b} Cs-IV. Acuostic modes are denoted with dashed lines.}
\label{Extended_fig1}
\end{figure*}

%\begin{appendices}

%\section{Section title of first appendix}\label{secA1}

%An appendix contains supplementary information that is not an essential part of the text itself but which may be helpful in providing a more comprehensive understanding of the research problem or it is information that is too cumbersome to be included in the body of the paper.




%\end{appendices}


\end{document}